# Spin driven emergent antiferromagnetism and metal insulator transition in nanoscale p-Si

Paul C Lou[1], and Sandeep Kumar[1,2*]

[1] Department of Mechanical Engineering, University of California, Riverside, CA

[2] Materials Science and Engineering Program, University of California, Riverside, CA

Abstract

The entanglement of the charge, spin and orbital degrees of freedom can give rise to emergent behavior especially in thin films, surfaces and interfaces. Often, materials that exhibit those properties require large spin orbit coupling. We hypothesize that the emergent behavior can also occur due to spin, electron and phonon interactions in widely studied simple materials such as Si. That is, large intrinsic spin-orbit coupling is not an essential requirement for emergent behavior. The central hypothesis is that when one of the specimen dimensions is of the same order (or smaller) as the spin diffusion length, then non-equilibrium spin accumulation due to spin injection or spin-Hall effect (SHE) will lead to emergent phase transformations in the non-ferromagnetic semiconductors. In this experimental work, we report spin mediated emergent antiferromagnetism and metal insulator transition in a Pd (1 nm)/$Ni_{81}Fe_{19}$ (25 nm)/MgO (1 nm)/p-Si (~400 nm) thin film specimen. The spin-Hall effect in p-Si, observed through Rashba spin-orbit coupling mediated spin-Hall magnetoresistance behavior, is proposed to cause the spin accumulation and resulting emergent behavior. The phase transition is discovered from the diverging behavior in longitudinal third harmonic voltage, which is related to the thermal conductivity and heat capacity.

Si is the apex semiconductor and an important material for spintronic application because of weak spin-orbit coupling and absence of spin scattering mechanisms[1]. Since spin-phonon interactions are the primary mechanism of spin relaxation in Si, we hypothesize that reduction of phonon population, occupation and mean-free-path can enhance the spin accumulation, allowing the manifestation of coherent spin states in p-Si. The site inversion asymmetry in lattice structure of Si has been proposed to cause hidden (or local) antiferromagnetic (AFM) exchange interaction[2, 3]. The hidden AFM interaction may be enhanced to strong AFM interactions with introduction of spin polarization, resulting in the spin mediated emergent behavior[4-6]. The emergent behavior is not intrinsic to the individual entities and can appear due to coupling across physical domains (electrical, thermal, magnetic, chemical etc.) leading to complex behavior as a collective. In this study, spin-charge and phonon coupling may give rise to emergent AFM behavior, which is not intrinsic to either p-Si or $Ni_{81}Fe_{19}$ layers. The spin mediated emergent AFM phase transition is considered as second order phase transformation, which can be discovered using thermal transport measurements[7-13]. The p-Si has been experimentally shown to exhibit inverse spin-Hall effect[14]. Hence, it should have spin-Hall effect (SHE) as well due to reciprocity. The spin accumulation due to SHE when absorbed at the ferromagnet/semiconductor interface will create spin polarization in the semiconductor. The proposed spin polarization mechanism is adapted from the observation of spin-Hall magnetoresistance (SMR)[15, 16] in ferromagnetic metal/heavy metal bilayers. The intrinsic spin Hall angle of p-Si is extremely small[14] ($10^{-4}$) and may not lead to observable SMR signal. But, the spin mediated thermal transport measurement can be used to study spin-phonon interactions and emergent behavior[17].

To enable emergent behavior, phonon mean-free-path must first be reduced. Studies have shown reduction in mean-free-path can be achieved with boundary scattering in in nanoscale and nanowires[18-21]. This is mimicked in our magneto-electro-thermal transport measurement setup having p-Si thickness similar to the spin diffusion length (~300 nm[22]). The micro-electro-mechanical systems (MEMS) experimental setup relies on spin-Hall effect to create spin polarization the p-Si layer as stated earlier. The experimental setup includes a freestanding Pd (1 nm)/ $Ni_{81}Fe_{19}$ (25 nm)/ MgO (1 nm)/ p-Si (400 nm) multilayer specimen. The MgO layer facilitate spin tunneling as well as act as a diffusion barrier. To observe the spin mediated behavior, we recorded the longitudinal $V_{1\omega}$ (electrical resistance), $V_{2\omega}$ (spin Seebeck effect (SSE), anomalous Nernst effect (ANE), tunneling anisotropic magnetoresistance (TAMR), magneto-thermopower (MTP))[23-28] and $V_{3\omega}$ (self-heating $3\omega$ method for thermal conductivity[29]) responses. The application of electrical current always creates a parabolic (approximately) longitudinal temperature gradient in the specimen[30, 31]. In the thin film specimens on substrate, the resulting in-plane temperature gradient is insignificant and can be neglected. But in the case of a freestanding specimen, the longitudinal temperature gradient can be used to measure the in-plane thermal transport behavior of the specimen. The self-heating $3\omega$ method utilizes a time-dependent current of frequency $\omega$ and amplitude $I_0$ in the specimen to both generate the temperature fluctuations and probe the thermal response. The technique relies on the solution of the one-dimensional heat conduction equation for the specimen, which is given by

$$\rho C_p \frac{\partial \theta(x,t)}{\partial t} = \kappa \frac{\partial^2 \theta(x,t)}{\partial x^2} + \frac{I_o{}^2 \sin^2 \omega t}{LS}(R_o + R'\theta(x,t)), \qquad (1)$$

where L and S are the length between the voltage contacts and the cross-sectional area of the specimen, respectively. $\rho$, $Cp$ and $\kappa$ are the density, specific heat and thermal conductivity in the material. $R_0$ is the initial electrical resistance of the specimen at temperature $T_o$. $R'$ is the temperature derivative of the resistance $R' = \left(\frac{dR}{dT}\right)_{T_o}$ at $T_o$. $\theta(x,t) = T(x,t) - T_o$ is the temporal ($t$) and spatial ($x$) dependent temperature change, as measured along the length of the specimen, which coincides with the heat flow direction.

$$V_{3\omega} \approx \frac{4 I^3 R_o R' L}{\pi^4 S \kappa \sqrt{1 + (2\omega\gamma)^2}} \qquad (2)$$

where $\gamma$ is the thermal time constant and is related with the heat capacity $\left(C_p = \frac{\pi^2 \gamma \kappa}{\rho L^2}\right)$. The $V_{3\omega}$ is a function of both thermal conductivity and heat capacity. The thermal conductivity can be expressed in terms of the third harmonic voltage $V_{3\omega}$ in the low frequency limit by

$$\kappa \approx \frac{4 I^3 R_o R' L}{\pi^4 V_{3\omega} S} \qquad (3)$$

We can infer that the heat capacity and thermal conductivity can be considered as a function of resistance and $V_{3\omega}$ response $\left(f\left(\kappa, C_p\right) = \frac{R}{V_{3\omega}}\right)$.

(Figure 1)

The devices are fabricated (Supplementary Figure S1) using micro/nanofabrication methods. To fabricate the experimental setup, we took a commercially available SOI wafer with a B-doped 2 μm thick device layer having resistivity of 0.001-0.005 $\Omega$ cm. We chemically etched the device layer of the SOI wafer to achieve the thickness closer to ~400 nm, near the spin diffusion length, by successively oxidizing and etching the wet thermal oxide using hydrofluoric (HF) acid. Using UV photolithography and deep reactive ion etching (DRIE), we etched the handle layer underneath the specimen region. Then we patterned and etched the front setup in the Si device layer using DRIE. We made the silicon structure freestanding using hydrofluoric acid vapor etch. In the next step, we removed the surface oxide by Ar milling for 15 minutes followed by a layer of 1 nm of MgO using RF sputtering. A layer of 25 nm $Ni_{81}Fe_{19}$/ 1nm Pd is, then, deposited on to the device using e-beam evaporation. The material deposition using evaporation leads to line of sight thin film deposition. The Pd layer is to protect the specimen from oxidation. The MgO layer is deposited as a tunneling barrier. The fabrication process is shown in supplementary Figure S1.

The magneto-electro-thermal transport measurements are carried out inside Quantum Design physical property measurement system (PPMS). The ac bias of 290µA at 5 Hz is applied across the outer electrodes using Keithley 6221 current source; and corresponding $R_{1\omega}$, $V_{2\omega}$ and $V_{3\omega}$ are recorded using SR-830 lock-in amplifiers. The responses are measured as function of magnetic field from 3T to -3T at various temperatures between 5 K and 300 K as shown in Figure 1 b-d. From the magneto-electro-thermal transport data, we observe a negative magnetoresistance (MR) of -1.5% at 300 K, which gradually increases to -3.1% at 5 K. The negative MR originate from the $Ni_{81}Fe_{19}$ layer. The MR behavior presented in Figure 1 b shows a knee at ~1.25 T, which corresponds to the saturation magnetization of $Ni_{81}Fe_{19}$. We also observe that the specimen resistance at 300 K is ~299 $\Omega$ and at 200 K it is 290 $\Omega$, which is a small change for 100 K temperature difference (Supplementary Figure S2). We then analyzed the $V_{2\omega}$ response behavior to uncover the potential contribution of the SSE, ANE, TAMR and MTP. The $V_{2\omega}$ response measured at 300 K is very large and shows a linear behavior as a function of applied current (Supplementary Figure S3) as opposed to quadratic ($I^2$) behavior expected for SSE, ANE, TAMR and MTP. This linear behavior can be attributed to the electric current being shunted across the bulk of the $Ni_{81}Fe_{19}$ and p-Si layers. The $V_{2\omega}$ response shows a magnetic field dependent behavior for an out of plane magnetic field (z-direction). The SSE and ANE should be absent for out of plane magnetic field since the temperature gradient (z-direction) across the thickness of the specimen will be parallel to the spin polarization and magnetization. This lead us to believe that SSE and ANE are not the primary cause of observed $V_{2\omega}$ response. The TAMR and MTP requires either heavy metal with large spin-orbit coupling or Rashba spin-orbit coupling (SOC) at the interface[26, 32]. The p-Si have insignificant intrinsic spin-orbit coupling but the $Ni_{81}Fe_{19}$/MgO/p-Si interface may lead to Rashba SOC essential for TAMR and MTP. The magnetic field dependent $V_{2\omega}$ response shows a saturation behavior at ~1.25 T corresponding to the saturation magnetization of $Ni_{81}Fe_{19}$ layer. The saturation behavior in $V_{2\omega}$ response can originate due to SHE mediated spin absorption/reflection at the interface. We propose that the spin polarization in the p-Si layer lead to spin-phonon interactions and resulting $V_{2\omega}$ response due to MTP observed in this study. The SSE may also contribute to spin-phonon interactions and resulting $V_{2\omega}$ response. The $V_{3\omega}$ response shows a magnetic field dependent behavior, and this can be interpreted as a spin influenced thermal transport in p-Si since p-Si is significantly more thermally conducting than $Ni_{81}Fe_{19}$. We estimate that the thermal resistance p-Si layer will be ~26 times (assuming a $\kappa_{p-Si}$ =30 W/mK) of $Ni_{81}Fe_{19}$ layer ($\kappa_{Ni_{81}Fe_{19}}$ = 21 W/mK[33]). In addition, the magnetic field dependent measurements of $V_{2\omega}$ and $V_{3\omega}$ responses show temperature dependent minima between 20 K and 50 K.

To uncover the insignificant temperature dependent change in resistance and to measure the $R'$ for thermal conductivity calculations, we acquired the $R_{DC}$ as a function of temperature from 350 K to 5 K at direct current of 10 µA to minimize the heating. The resistance is measured using a Keithley 6221 current source and 2182A nanovoltmeter. The temperature dependent resistance behavior shows a rapid increase and then continuous decrease after ~250 K as shown in Figure 2 a. We then heated the specimen from 5 K with an applied magnetic field of 1.25 T and stopped at 140 K. At this point, we cooled the specimen again in the presence of applied magnetic field of 1.25 T. We observe that the field-cooling (FC) curve starts to separate from field-heating (FH) curve and both meet around 20 K. We carried out the FH again starting from 20 K to 300 K. And the FH behavior follows the first curve indicating a hysteretic behavior. Although the current density in this case is ~2.64x10$^2$ A/cm$^2$ but hysteretic behavior may originate from the thermal drift because the specimen is freestanding. In addition, the box oxide layer of the SOI wafer is 1 µm thick may induce thermal lag in the measurements. The observed diverging resistance behavior shown in Figure 2a can be interpreted as metal-insulator transition (MIT) in p-Si layer due to either ferromagnetic proximity or spin accumulation (SHE) leading to shunting of the electric current across $Ni_{81}Fe_{19}$ layer.

To understand the origin of observed behavior, we acquired $V_{2\omega}$ response as a function of temperature under zero applied magnetic field. We cooled the specimen at 0.3 K/min from 400 K to 200 K followed by heating from 200 K to 300 K. We cooled the specimen again from 300 K to 5 K followed by heating it to 300 K. The measured $V_{2\omega}$ response is presented in Supplementary Figure S4 a. We observe an inflection point in the $V_{2\omega}$ response at ~360 K, which may indicate advent of spin dependent behavior. Next, we acquired the $R_{1\omega}$ and $V_{3\omega}$ as a function of temperature while the specimen was cooled from 350 K to 5 K at 0.3 K/min for $I_{rms}$ of 290μA. The measured $V_{3\omega}$ data shows an inflection point at 259 K as shown in Figure 2 b. The $R'$ has a peak at ~268 K, making the thermal conductivity undefined around the peak due to zero slope. The $V_{3\omega}$ is a function of thermal conductivity only in the case of low frequency, and the observed behavior may violate the low frequency assumption. In that case, $V_{3\omega}$ will be a function of both thermal conductivity and heat capacity (through thermal time constant- γ). We plotted $f(\kappa, C_p) = \frac{R}{V_{3\omega}}$ as a function of temperature to understand the thermal property behavior in the absence of valid $R'$. We observe a sharp peak in this data and diverging behavior in thermal transport as shown in Figure 2 c. We observe that the $\frac{R}{V_{3\omega}}$ increases from 0.725 Ω/μV at 300 K to 764.2 Ω/μV at 259 K. Since, the second order phase transformations are characterized from singularities or discontinuities in the temperature dependent heat capacity measurements, the diverging behavior in $\frac{R}{V_{3\omega}}$ can be considered a second order phase transformation. In this study, we have used $V_{3\omega}$ response to uncover the phase transition behavior instead, which is a function of thermal conductivity and heat capacity especially near the phase transition. Then, we heated the specimen under an applied magnetic field of 1.25 T at 0.3 K/min. The field dependent heating shows a shift in inflection point in $V_{3\omega}$ to 267 K, which is also attributed to the thermal drift.

(Figure 2)

From the temperature dependent study, we propose that the second order AFM phase transformation is the underlying cause of inflection point observed in the temperature dependent $V_{3\omega}$ measurement. To understand the effect of applied magnetic field on the phase transformation, we carried out temperature dependent $V_{1\omega}$, $V_{2\omega}$ and $V_{3\omega}$ measurement for an applied magnetic field of 14 T as shown in Figure 2 d and Supplementary Figure S4 b-c. The qualitative behavior for $R_{1\omega}$ as a function of temperature for applied magnetic field of 14 T is similar to the $R_{DC}$ data presented earlier, except the resistance values are lower due to negative magnetoresistance and peak has shifted toward higher temperature. The $V_{2\omega}$ response shows minimal field dependent changes in the behavior as compared with zero field measurement shown in Supplementary Figure S4 b. The inflection point for $V_{3\omega}$ response is shifted ~250 K due to applied magnetic field from 259 K at zero field. The magnetic field has a measurable but small effect on the phase transition behavior.

The diverging resistance behavior as a function of temperature, shown in Figure 2 a and 2 d, can be considered a metal-to-insulator transition (MIT). To uncover the origin of it, we measured the resistance of a p-Si specimen (having similar resistivity) and $Ni_{81}Fe_{19}$ specimens as a function of temperature as shown in Supplementary Figure S5-S6 respectively. The p-Si specimen shows a semiconductor behavior; the peak in electrical resistance occurs below 50 K and resistance does not increase significantly until ~200 K. The p-Si layer loses the metallic behavior (dopants) during oxidation-based chemical thinning methods utilized in the present work (methods), which may be the reason for semiconducting behavior. The resistance of the $Ni_{81}Fe_{19}$ layer is estimated to be ~365 $\Omega$ and ~305.6 $\Omega$ at 350 K and 267 K respectively (Supplementary Figure S6).

The resistance at 350 K is estimated using the linear temperature coefficient of resistance. It needs to be stressed that the dimensions of control $Ni_{81}Fe_{19}$ specimen are different from the $Ni_{81}Fe_{19}$ layer (Supplementary Figure S6). We can model the specimen having $Ni_{81}Fe_{19}$ and p-Si layers as two resistors in parallel. Based on change in resistance of $Ni_{81}Fe_{19}$ layer, we estimated that the resistance of p-Si will change from ~803.7 $\Omega$ at 350 K to ~13893 $\Omega$ at 267 K. Such large change in the p-Si resistance can only occur due to MIT as hypothesized. The resistances are estimated using the $R_{1\omega}$ data for the multilayer specimen. We propose that the spin accumulation in p-Si may induce an emergent ferromagnetic or AFM phase transition. The temperature dependent measurement of $V_{3\omega}$ response at 14 T (Supplementary Figure S4c) suggests AFM phase transition and not ferromagnetic transition; since the transition behavior is weakly dependent on the applied magnetic field. We propose that the AFM spin-spin interactions lead to a gradual transition from conductor to insulating state as hypothesized in this work.

(Figure 3)

To replicate the experimental results, we carried out the temperature dependent measurement on another device, which has significantly lower resistance at room temperature indicating that the p-Si layer has a large charge carrier density (~10 times) relative to the first device. The resistance of the layered thin film specimen at 400 K is ~98.24 $\Omega$. Using a parallel resistor configuration, we estimate the resistance of p-Si layer to be ~150 $\Omega$, which is an order of magnitude lower than the first device. The data from the second device is shown in Figure 3 a-d. The data is acquired at cooling and heating rate of 0.2 K/min, which may reduce the thermal drift. In this device, we observe the proposed AFM transition at ~312 K, which corresponds to $V_{3\omega}$ magnitude of zero as shown in Figure 3 c and corresponding peak in $\frac{R}{V_{3\omega}}$ as shown in Figure 3 d. We also observe two additional transitions at ~236 K and ~50 K. The emergent AFM is not intrinsic to p-Si and multiple AFM states may exist at different temperature and charge carrier concentrations. We propose that the second transition at ~236 K is transition from one emergent AFM state to another AFM state. At low temperatures, the AFM interactions cause a MIT[34, 35], and resistance goes from 110 $\Omega$ to ~260 $\Omega$ as shown in Figure 3 a and 3 a-inset. The resistance (~260 $\Omega$) after MIT corresponds to the $Ni_{81}Fe_{19}$ layer only. Based on diverging behavior in $\frac{R}{V_{3\omega}}$, we can confirm that the observed transition in current study is a second order transition and not a structural. Earlier, we proposed that the $V_{2\omega}$ originated from the absorption of spin current from SHE in p-Si to $Ni_{81}Fe_{19}$ layer. After MIT, the $V_{2\omega}$ response should go to zero since SHE in p-Si will cease to exist in the absence of charge current across the p-Si layer. The temperature dependent $V_{2\omega}$ response clearly supports our argument as shown in Figure 3 b. In addition, we also observe a sharp drop in $V_{3\omega}$ response as well due to MIT from 2340 $\mu V$ to ~185 $\mu V$. The $V_{3\omega}$ response is inversely related with the thermal transport behavior. The drop in $V_{3\omega}$ due to MIT may signify recovery of large phononic thermal transport in p-Si, which is suppressed due to spin polarization before MIT. In addition, we also observe that the phase transition behavior is a function of doping level in p-Si. At low charge carrier density, the AFM transition is observed at ~259 K (from $R/V_{3\omega}$ response) while at higher charge carrier density the transition occurs at ~312 K. We propose that the observed MIT at higher temperature is Anderson disorder transition[34, 36].

The observed transition behavior in this study is attributed to the spin accumulation due to SHE. The spin accumulation should be a function of applied current density. Hence, the transition behavior should be a function of current density. To prove our hypothesis, we measured the $R_{1\omega}$, $V_{2\omega}$ and $V_{3\omega}$ as a function of temperature (at 0.2 K/min) for different applied electric currents (400 µA, 500 µA, 750 µA and 1 mA) as shown in Figure 4. The Joule heating should lower the transition temperature (spin relaxation) and electrical resistance after transition whereas enhanced polarization due to SHE may increase the transition temperature. The resistance corresponding to the peak reduces as the current is increased, which is inferred as effect of Joule heating. But, the transition temperature (deduced from the $\frac{R}{V_{3\omega}}$) increases as a function of applied electrical current as shown in Figure 4 a, c. The transition temperature is observed at 259 K for 290 µA of electrical current as stated earlier. The transition temperature is observed at 265 K, 270 K, 282 K and 288 K for applied current of 400 µA, 500 µA, 750 µA and 1 mA respectively. In addition, a second transition emerges for the applied current of 750µA and 1 mA, which may be inferred as transition from one AFM state to another and may be a precursor to the Anderson transition observed in the second device. These measurements demonstrate beyond reasonable doubt that the observed behavior is transport mediated.

(Figure 4)

To further support our argument and have a direct proof of the SHE, we measured the magnetoresistance and $V_{2\omega}$ response as a function of angular rotation of the constant magnetic field in the yz-plane at 350 K (before transition) and 200 K (after transition). This measurement allows us to identify the spin Hall magnetoresistance (SMR), anomalous magnetoresistance (AMR), SSE, ANE, TAMR and MTP. The magnetoresistance measurement at 350 K shows a response, which can be considered a combination of $\sin^2 \phi_{zy}$ and $\cos \phi_{zy}$ as shown in Figure 5 a and Supplementary Figure S7 a. The dominant $\sin^2 \phi_{zy}$ response originate from the SMR and this response disappears at 200 K as shown in Supplementary Figure S7 b. AT 200 K, the resistance response is entirely due to AMR of $Ni_{81}Fe_{19}$ layer, which is confirmed from the $Ni_{81}Fe_{19}$ control specimen. From the $V_{2\omega}$ response shown in Supplementary Figure S7 c-d, we can eliminate the existence of ANE and SSE since we do not observe the sine dependence. At 350 K, we observe a combination of $\sin^2 \phi_{zy}$ and $\cos \phi_{zy}$ behavior in the $V_{2\omega}$ response. We relate the $\cos \phi_{zy}$ behavior to spin absorption due to SHE and $\sin^2 \phi_{zy}$ may originate from the MTP as proposed earlier. The $\cos \phi_{zy}$ behavior suggest spin current absorption when the magnetization is in z-dir and reflection when the magnetization is in y-dir. The $\cos \phi_{zy}$ behavior disappears at 200 K due to the phase transition and only $\sin^2 \phi_{zy}$ behavior due to MTP is observed. The SMR behavior is surprising since the spin Hall angle of p-Si has been reported to be insignificant. The p-Si layer in the multilayer specimen in this study is metallic at 350 K. Hence, to calculate the $\theta_{SH}$, we utilize the SMR equations for bimetallic[37] specimen. The MgO layer in the present study is insulating and we can neglect it for SMR calculations.

$$\frac{\Delta R_{xx}^{SMR}}{R_{xx}^0} \sim -\theta_{SH}^2 \frac{\lambda_N}{d} \frac{tanh^2\left(\frac{d}{2\lambda_N}\right)}{1+\xi} \left[ \frac{g_R}{1+g_R coth\left(\frac{d}{\lambda_N}\right)} + \frac{g_F}{1+g_F coth\left(\frac{d}{\lambda_F}\right)} \right] \qquad (4)$$

where $g_R \equiv 2\rho_N \lambda_N Re[G_{MIX}]$ and $g_F \equiv \frac{(1-P^2)\rho_N \lambda_N}{\rho_F \lambda_F coth\left(\frac{t_F}{\lambda_F}\right)}$.

We use the following values to calculate the SMR[38]: $\rho_N = 3.17 X 10^{-5} \, \Omega m, \lambda_N = 230 \, nm, Re[G_{MIX}] = 10^{19}\Omega^{-1}m^{-2}, P = 0.7, \ \rho_F = 3.97 X 10^{-7}\Omega m, \lambda_F = 4 \, nm, t_F = 25 \, nm \ and \ d = 400 \, nm$.

From these values, we realize that $1 \ll g_R \coth\left(\frac{d}{\lambda_N}\right)$ and $1 \ll g_F \coth\left(\frac{d}{\lambda_N}\right)$. This simplifies the relationship to:

$$\frac{\Delta R_{xx}^{SMR}}{R_{xx}^0} \sim -\theta_{SH}^2 \frac{\lambda_N}{d} \frac{2*tanh^2\left(\frac{d}{2\lambda_N}\right)}{(1+\xi)\coth\left(\frac{d}{\lambda_N}\right)} \tag{5}$$

For $\frac{\Delta R_{xx}^{SMR}}{R_{xx}^0} = 0.002$, we calculate $\theta_{SH} \sim 0.05$.

We calculate the spin-Hall angle ($\theta_{SH}$) to be 0.05, which is significantly larger than the $\theta_{SH}=10^{-4}$ reported for p-Si[14] and it is of the same order as Pt[39]. This lead us to believe that the ISHE essential for SMR cannot arise intrinsically in p-Si due to small intrinsic SOC (0.044 eV). The Rashba SOC[5, 6] due to broken structural symmetry at the MgO/p-Si interface can lead to SMR behavior as shown in Figure 5 c. The Rashba SOC is believed to occur in two-dimension electron gas (2DEG)[5]. Since most of the applied current is carried by the bulk $Ni_{81}Fe_{19}$ and p-Si layers, the observed SMR behavior is surprising and cannot originate only from the interfacial 2DEG. We hypothesize that the p-Si layer exhibits Rashba SOC leading to both SHE and ISHE. The structural inversion asymmetry essential for Rashba effect occurs in the p-Si layer across the whole thickness and not only at the interface. The SOC due to structural inversion asymmetry in Si metal-oxide semiconductor field effect transistor (FET) leads to suppression of spin resonance [40], which supports the proposed Rashba SOC in this work. But, the SOC in FET structures is weak due to insignificant intrinsic SOC in Si. Whereas in this study, $Ni_{81}Fe_{19}$ have relatively large intrinsic SOC[41, 42], which may give rise to the strong Rashba SOC due to proximity effect[43]. In addition, recent observation of the strong Rashba spin split states at Bi/Si(111) interface[44] also corroborates our hypothesis of Rashba SOC in the Pd/$Ni_{81}Fe_{19}$/ MgO/p-Si multilayer specimen. The Rashba SOC can also give rise to the MIT observed in this study[45, 46]. It needs to be stressed that the $Ni_{81}Fe_{19}$ thin films do not show MIT behavior and it only arises in p-Si due to proximity induced Rashba SOC. The interfacial SOC should also contribute towards SSE, which we do not observe. The absence of SSE can be attributed to the $T_{magnon}$ ($Ni_{81}Fe_{19}$) < $T_{phonon}$ (p-Si)[47] causing the spin backflow to $Ni_{81}Fe_{19}$ to be larger than the spin-Seebeck tunneling, which is supported by the $cos\,\phi_{zy}$ behavior observed in the second harmonic response at 350 K. The SMR behavior is supported by the angular field rotation measurement in zx and xy-planes as shown in Figure 5 a.

To further support our argument, we carried out the magnetic characterization using Quantum Design magnetic property measurement system (MPMS) as show in Supplementary Figure S8 a-b. We do not observe phase transition in temperature dependent magnetic moment measurement at 20 Oe as shown in Supplementary Figure S8 a. The ferromagnetic interactions can be attributed to the proximity effect to the ferromagnetic layer. But we observe insignificant exchange bias (~5 Oe) in the magnetic hysteresis measured at 300 K, 168 K and 5 K as shown in Supplementary Figure S8 b, which is attributed to the remnant magnetization in the superconducting magnet coils. The magnetic characterization eliminates Ni or Fe diffusion from $Ni_{81}Fe_{19}$ layer into p-Si being the cause of observed behavior. These measurements support that the observed behavior is transport-mediated phenomenon and not due to Fe/Ni doping or induced only by proximity effect. To verify our argument, we fabricated experimental MEMS device having a Hall bar structure to measure the change in anomalous Hall effect (AHE). We measured the transverse resistance at 350 K and 200 K as a function of magnetic field (out of plane) from 14 T to -14 T. The 350 K and 200 K lie on the either side of the phase transition temperature. The AHE measurements at 350 K and 200 K exhibits a reduction in anomalous Hall resistance ($R_{AH}$) as shown in Figure 5 b. Using line fit we identified the intercepts and calculated the $R_{AH}$. The $R_{AH}$ reduces from 71.91 m$\Omega$ at 350 K to 54.34 m$\Omega$ at 200 K. From the magnetic moment measurements presented in Supplementary Figure S8 a, we know that $R_{AHE}$ should increase as the temperature is reduced since the magnetic moment increases. The observed $R_{AHE}$ behavior may arise if the spin accumulation in the p-Si leads to canted AFM states and the net moment is opposite of $Ni_{81}Fe_{19}$ magnetization. From the comprehensive experimental measurements presented in this work, we confirm that the SHE induces spin polarization leads to the transition from weakly local AFM to strong global AFM behavior in p-Si. The schematic of the AFM phase transition mechanism is shown in Figure 5 d. The SHE is essential for this behavior, and hence magnetic characterization does not exhibit phase transition behavior. (Figure 5)

In conclusion, we report experimental proof of emergent antiferromagnetic and metal insulator phase transformation in nanoscale p-Si thin films. High temperature antiferromagnetic phase transformation is discovered through magneto-electro-thermal transport measurements. The phase transition is confirmed from the diverging behavior in resistance and $V_{3\omega}$ measurement. The SHE induces spin polarization and the lattice site inversion asymmetry in diamond cubic Si is proposed to be the underlying cause of emergent antiferromagnetic behavior in p-Si. The SHE is confirmed by SMR measurement and Rashba spin-orbit coupling is the mechanism of SHE and ISHE. The spin mediated emergent phase transition is a function of charge carrier concentration in p-Si. At low carrier density, p-Si behaves as a semiconductor and AFM interactions lead to AFM phase transformation. At high charge carrier concentration, the AFM interactions p-Si shows an AFM transition at high temperatures and a distinct MIT transition. The Si is not only the apex semiconductor but also one of the most widely characterized materials as well. Si thin films are widely used for semiconductor electronics, MEMS sensors and actuators for various applications. The observed emergent antiferromagnetic behavior may lay the foundation of Si spintronics and may change every field involving Si thin films. These experiments also present potential electric control of magnetic behavior using simple semiconductor electronics physics. The observed large change in resistance and doping dependence of phase transformation encourages development of antiferromagnetic and phase change spintronics devices.

**Acknowledgement**

We thank Prof. Ward Beyermann (UCR) for discussions and inputs.

List of figures

Figure 1. a. False color scanning electron micrograph showing the device structure, b. magnetoresistance behavior of the specimen, c. the magnetic field dependent $V_{2\omega}$ at 300 K and inset showing the $V_{2\omega}$ at different temperatures, and d. the magnetic field dependent $V_{3\omega}$.

Figure 2. a. $R_{DC}$ as a function of temperature, b. $V_{3\omega}$ as a function of temperature, c. the $\frac{R}{V_{3\omega}}$ as a function of temperature showing the peak corresponding to phase transformation, and d. the $R_{1\omega}$ as a function of temperature for out of plane magnetic field of 14 T. Arrows show the direction of temperature change.

Figure 3. a. $R_{1\omega}$ as a function of temperature showing MIT, b. the $V_{2\omega}$ as a function of temperature at zero magnetic field, c. the $V_{3\omega}$ as a function of temperature, and d. the $\frac{R}{V_{3\omega}}$ as a function of temperature showing the peak corresponding to phase transformations.

Figure 4. a. $R_{1\omega}$ as a function of temperature and b. the $V_{3\omega}$ response as a function of temperature, c. the $\frac{R}{V_{3\omega}}$ as a function of temperature showing the peak corresponding to phase transformations and d. the $V_{2\omega}$ as a function of temperature for applied electric currents 400 µA, 500 µA, 750 µA and 1 mA.

Figure 5. a. The angular field rotation in xy, zx and zy plane showing the SMR behavior, b. The Hall resistance measurement at 350 K and 200 K for an out of plane magnetic field showing reduction in anomalous Hall resistance and c. the schematic showing the SMR behavior due to intrinsic SHE and Rashba SOC and d. the schematic displaying the proposed mechanism of observe emergent antiferromagnetic behavior.

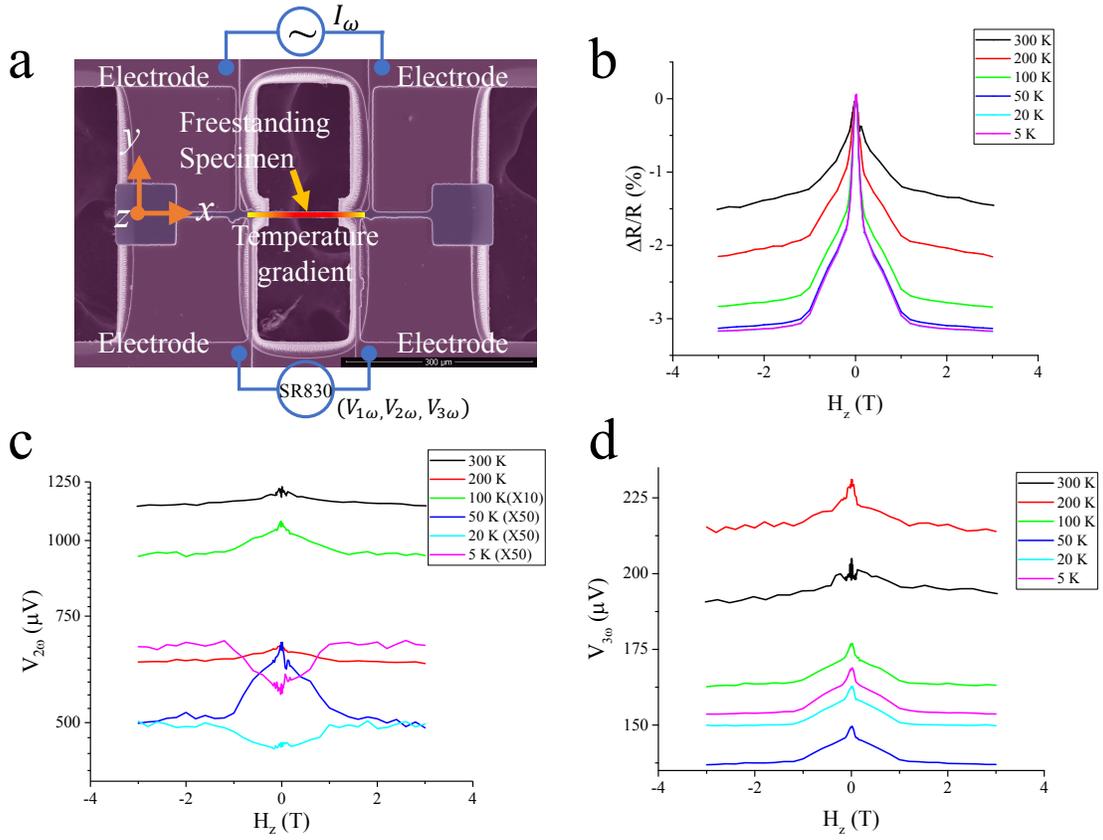

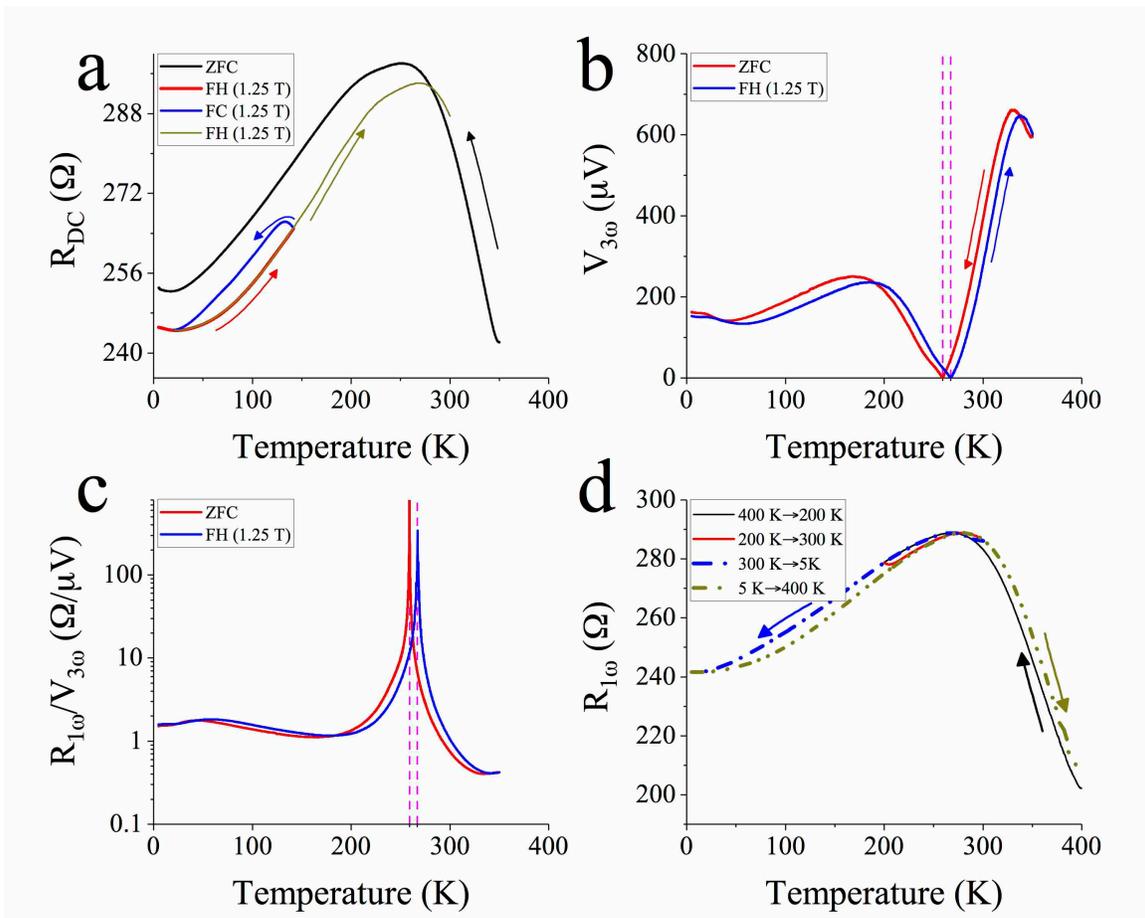

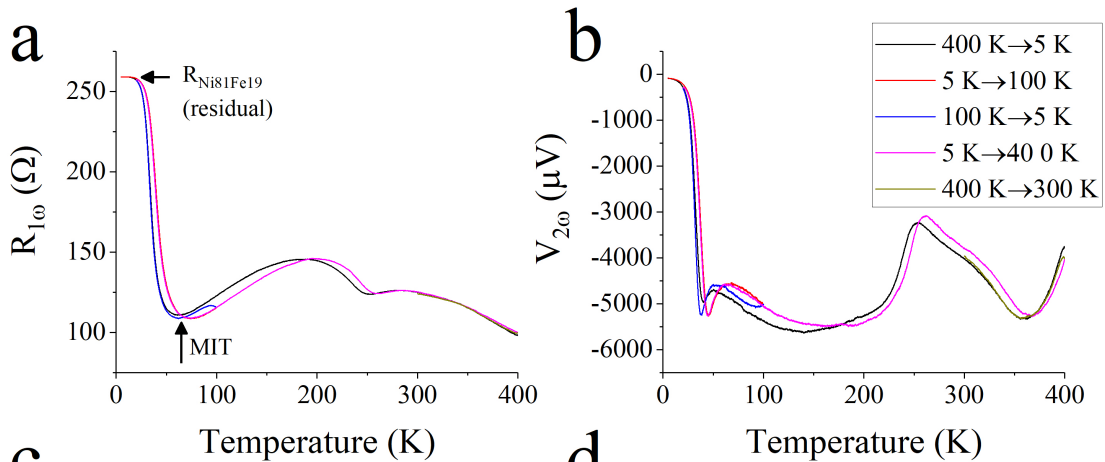

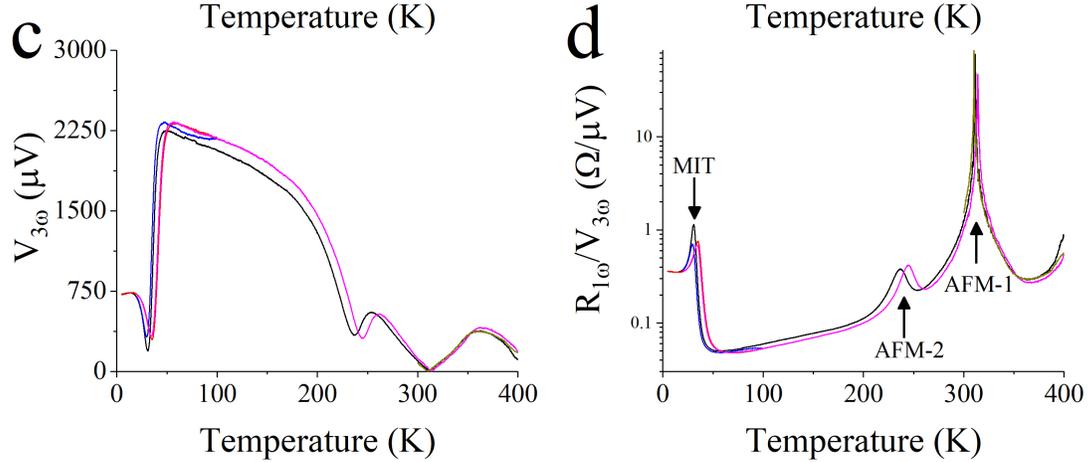

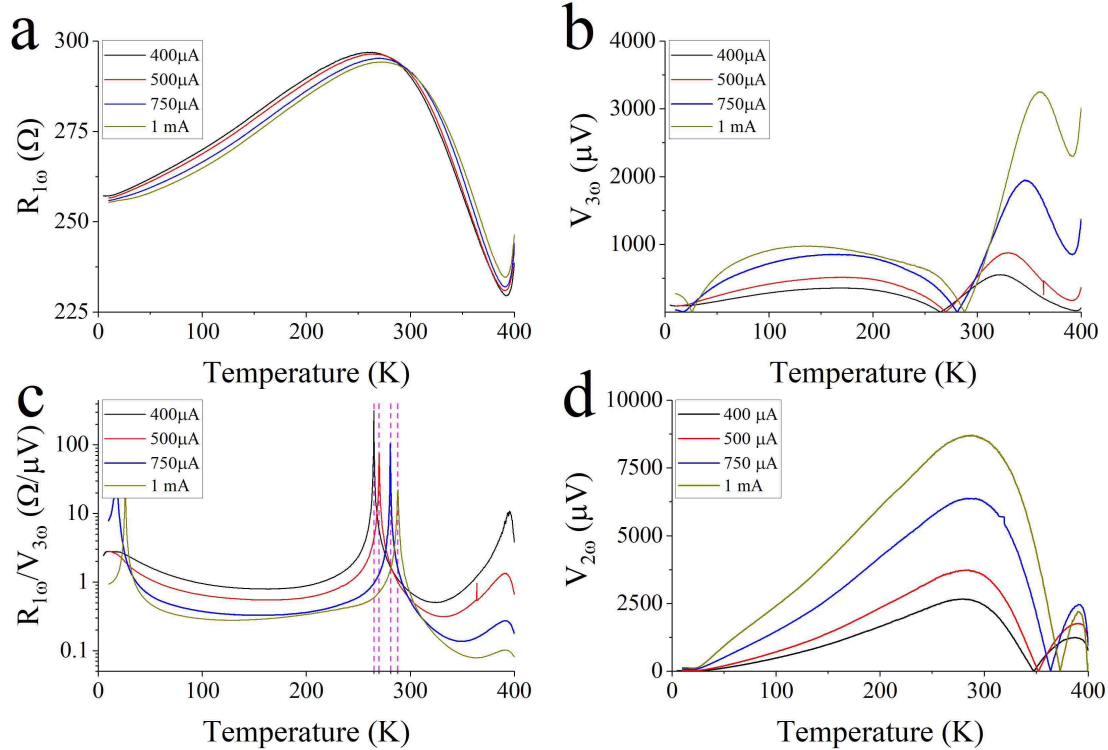

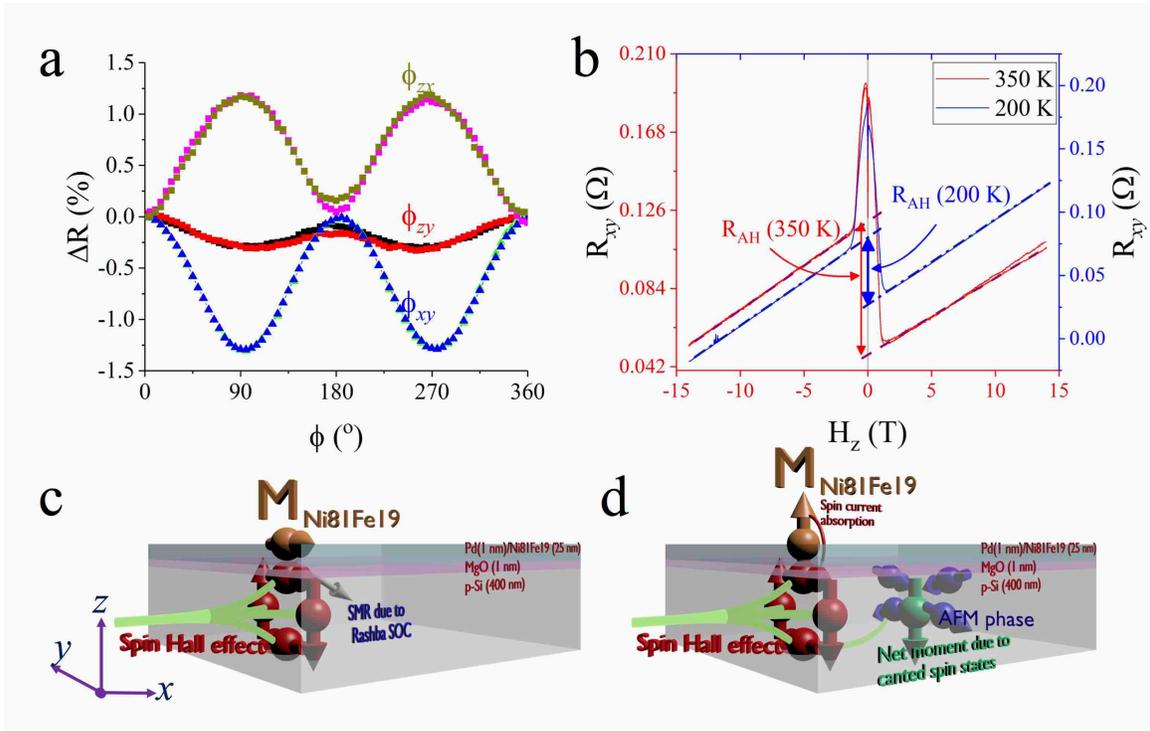

**Supplementary Figures**

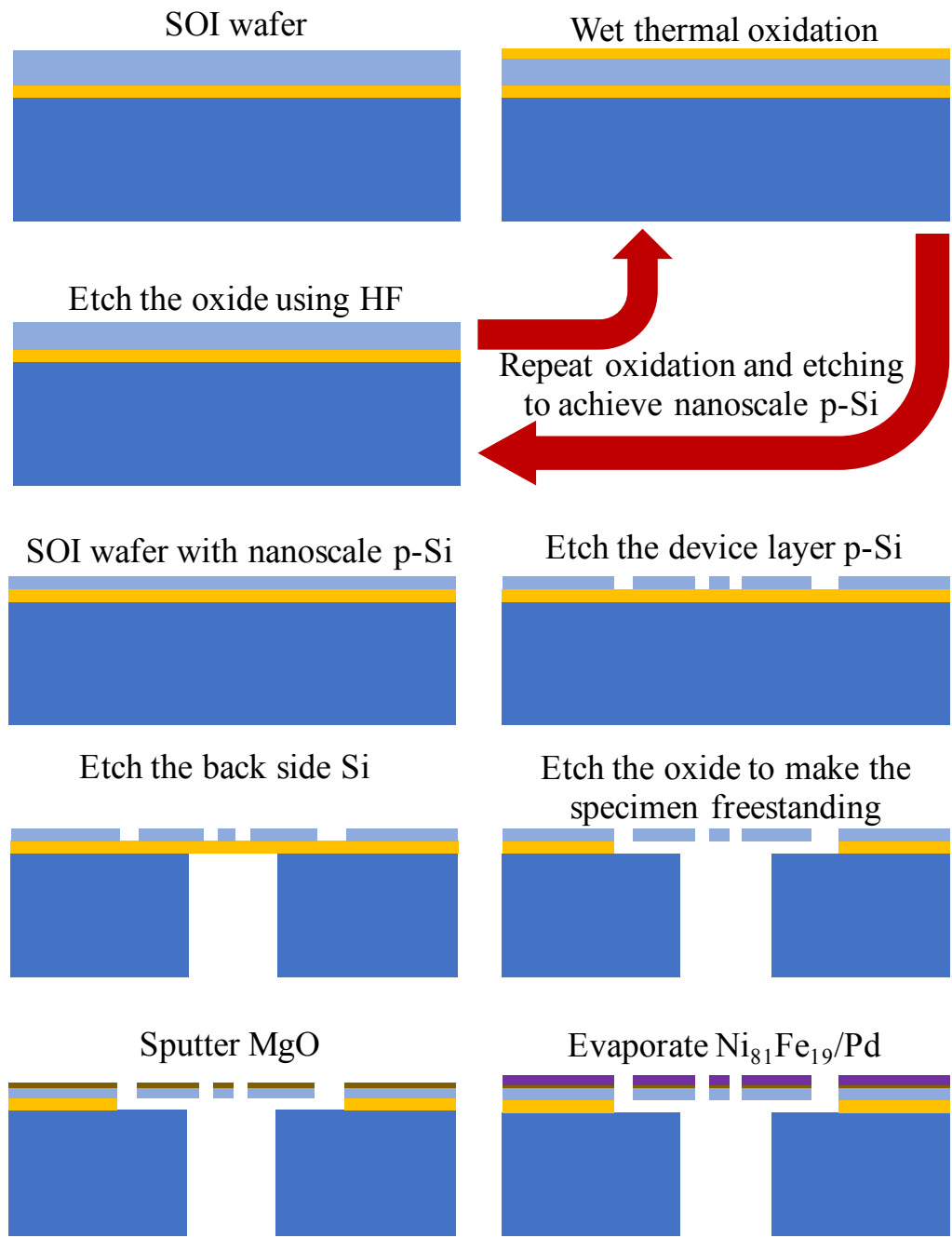

SOI wafer

Wet thermal oxidation

Etch the oxide using HF

Repeat oxidation and etching to achieve nanoscale p-Si

SOI wafer with nanoscale p-Si

Etch the device layer p-Si

Etch the back side Si

Etch the oxide to make the specimen freestanding

Sputter MgO

Evaporate $Ni_{81}Fe_{19}/Pd$

Supplementary Figure S1. The experimental device fabrication process.

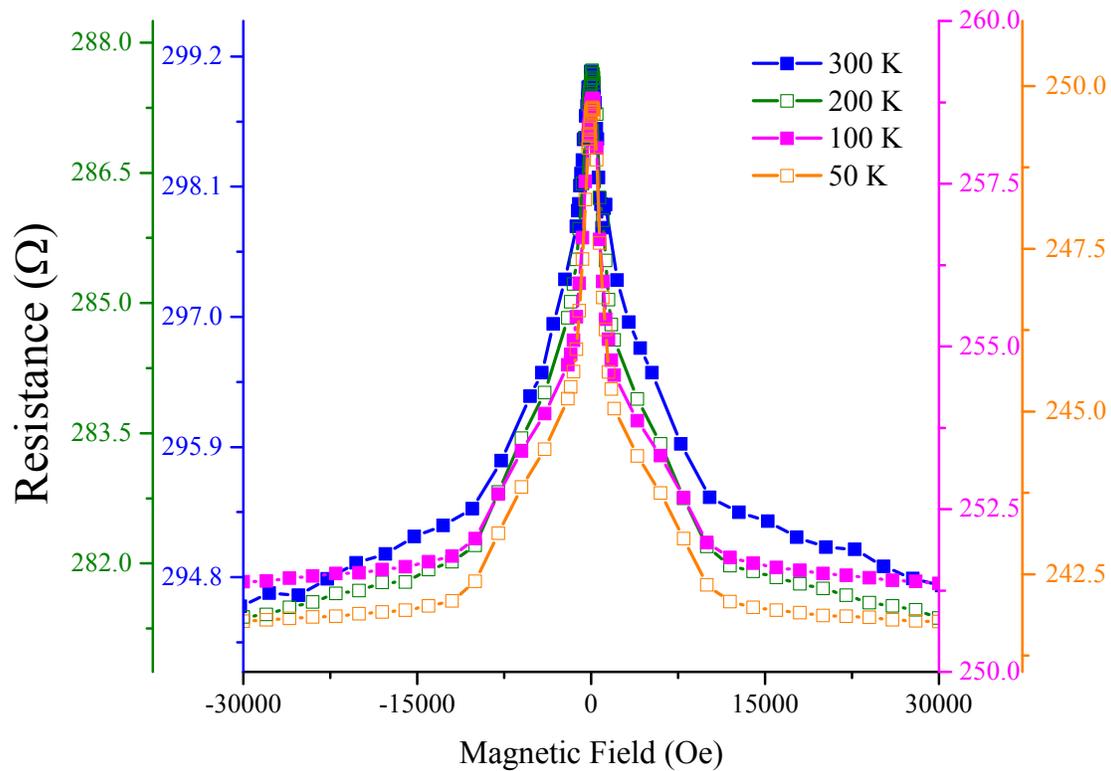

Supplementary Figure S2. The magnetoresistance measurement as a function of temperature.

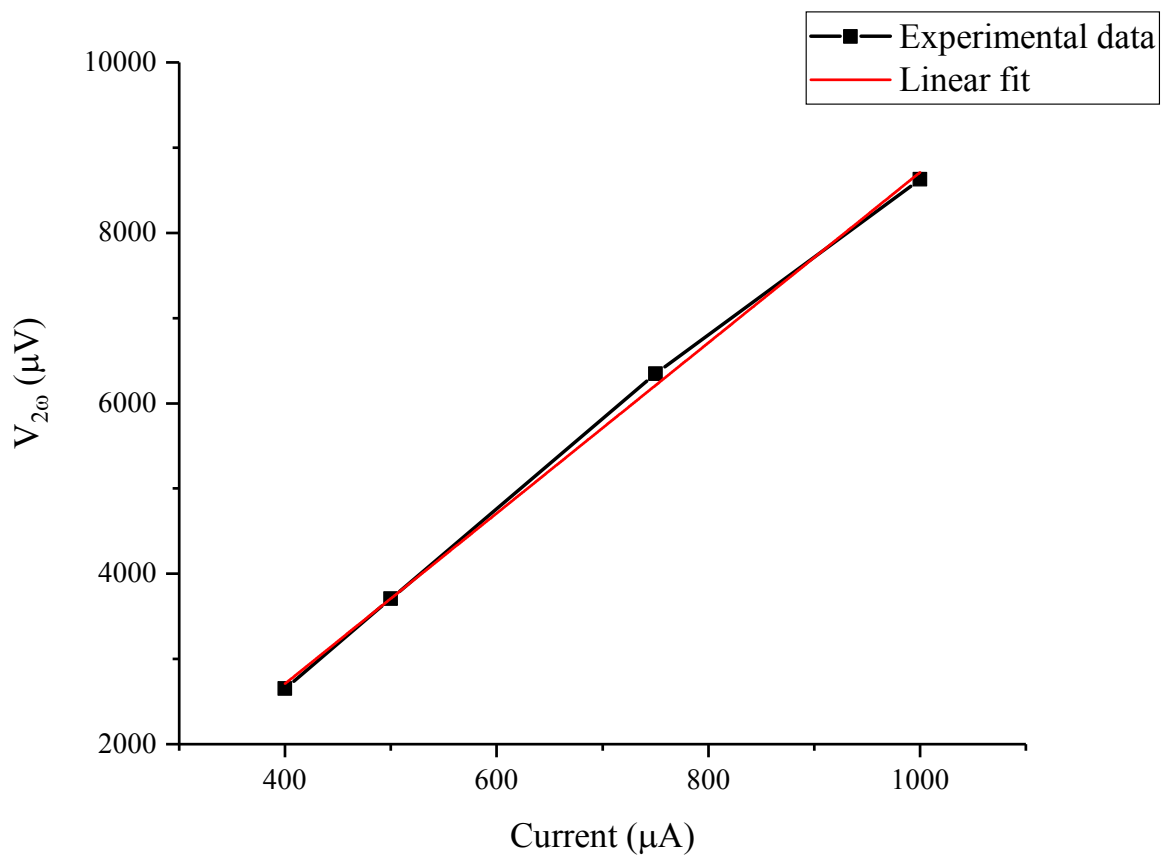

Supplementary Figure S3. The $V_{2\omega}$ response showing a linear behavior as a function of applied current.

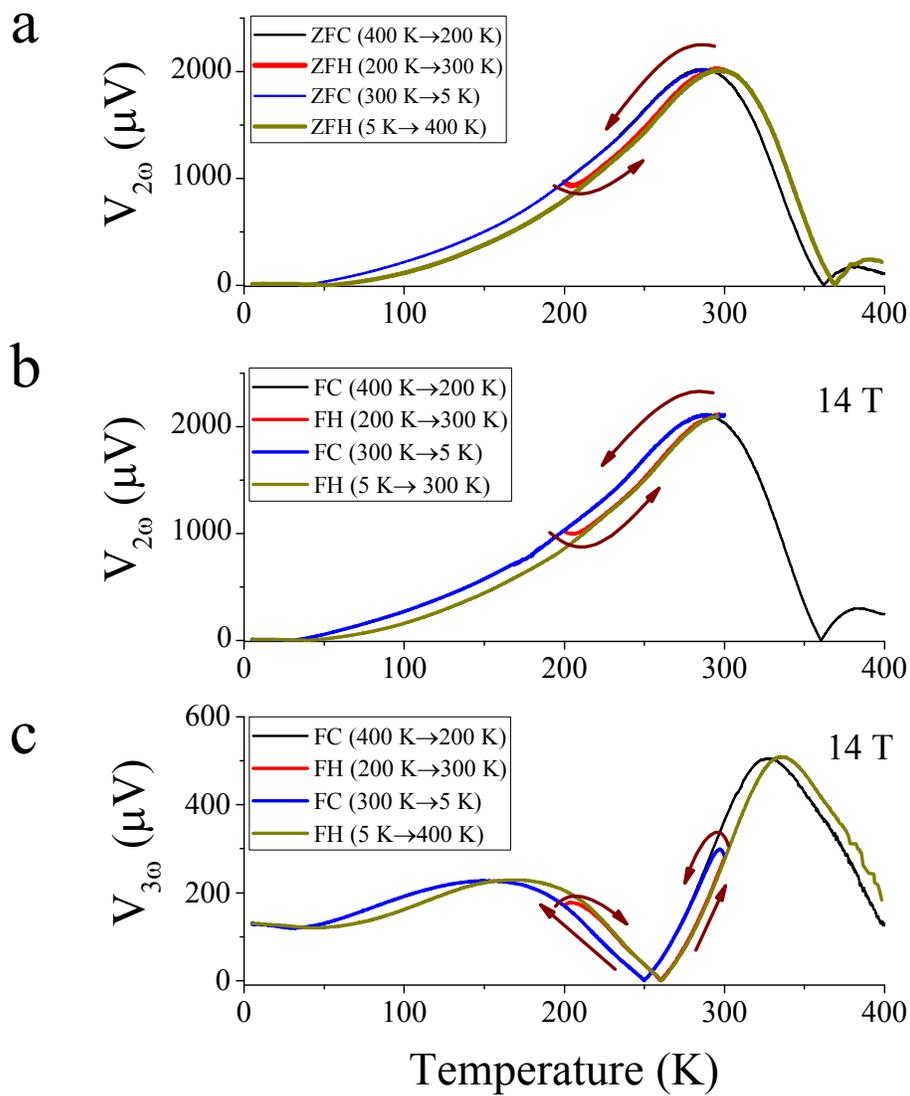

Supplementary Figure S4. a. the $V_{2\omega}$ response as a function of temperature at zero magnetic field, and b-c. the $V_{2\omega}$ and $V_{3\omega}$ responses as a function of temperature for out of plane magnetic field of 14 T. Arrows show the direction of temperature change.

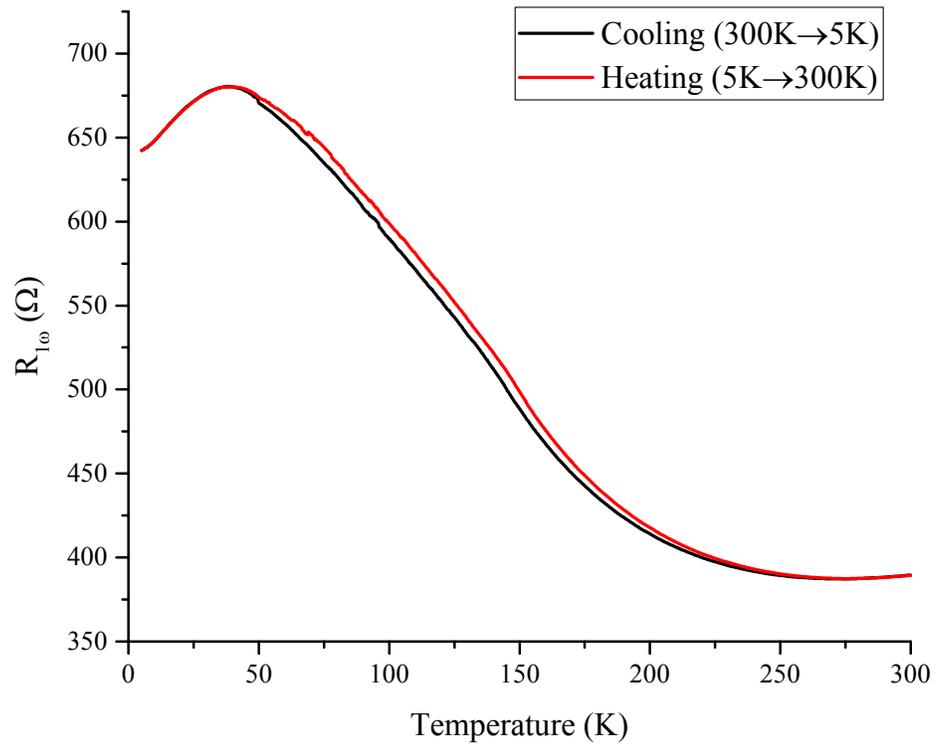

Supplementary Figure S5. The temperature dependent longitudinal resistance of p-Si specimen having length ~40 μm, width ~19 μm and thickness of 400 nm.

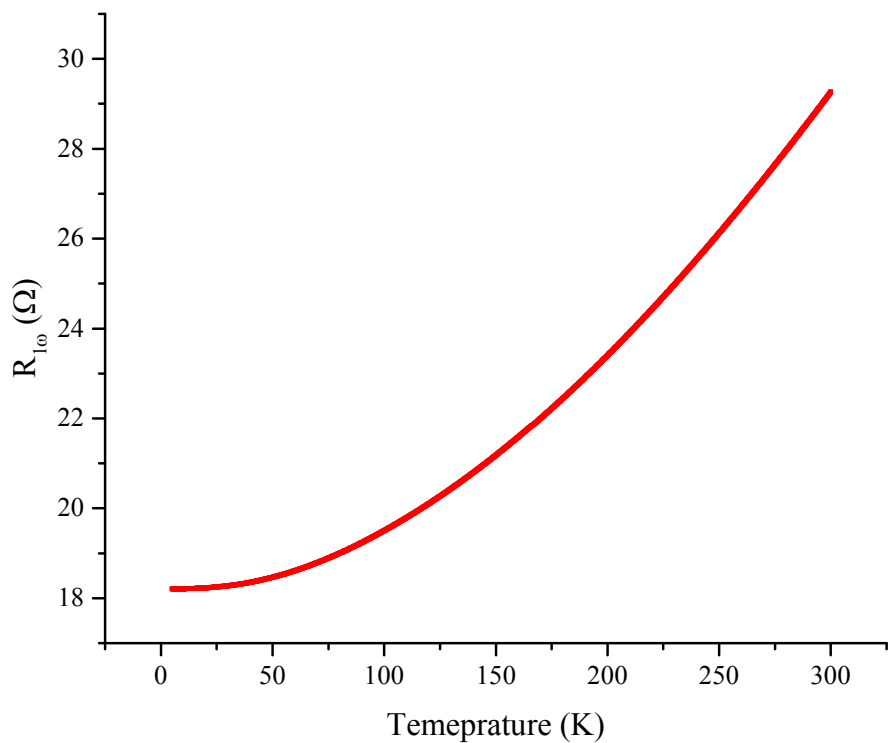

Supplementary Figure S6. The temperature dependent longitudinal resistance of Ni$_{81}$Fe$_{19}$ specimen having length ~40 μm, width ~19 μm and thickness of 25 nm.

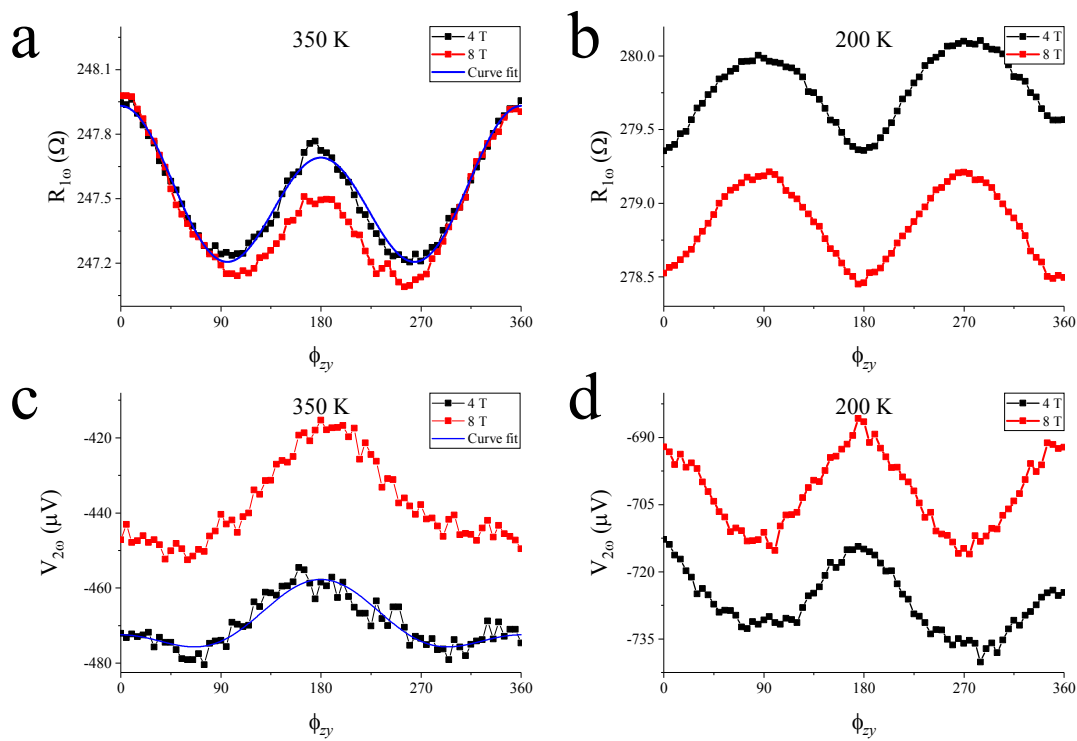

Supplementary Figure S7. The angular field rotation in zy plane showing a. resistance at 350 K, b. resistance at 200 K, c. the $V_{2\omega}$ response at 350 K and d. the $V_{2\omega}$ response at 200 K.

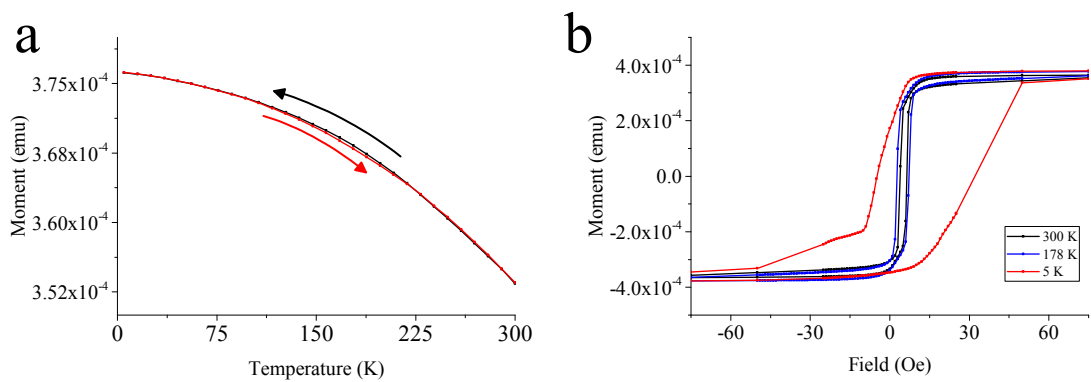

Supplementary Figure S8 a. The magnetic moment as a function of temperature for an applied magnetic field of 20 Oe and b. the magnetic hysteresis at 300 K, 178 K and 5 K.